\documentclass{article}
\usepackage{spconf,amsmath,graphicx}
\usepackage{amsfonts,amssymb}
\usepackage{multirow}
\usepackage{graphicx}
\usepackage{caption}
\usepackage{subcaption}
\usepackage{flushend}
\usepackage{bm}
\usepackage{booktabs}

\title{DeepChorus: A Hybrid Model of Multi-scale Convolution and Self-attention for Chorus Detection}
\name{Qiqi He $^1$, Xiaoheng Sun $^1$, Yi Yu $^2$ and Wei Li $^{1,3}$}
\address{$^1$ School of Computer Science and Technology, Fudan University, Shanghai, China \\
	$^2$ National Institute of Informatics (NII), Tokyo, Japan\\
	$^3$ Shanghai Key Laboratory of Intelligent Information Processing, Fudan University, Shanghai, China\\
\tt\small \{heqq20, 19210240112, weili-fudan\}@fudan.edu.cn, yiyu@nii.ac.jp
\vspace{-.3cm}}

\begin{document}
	\ninept
	\maketitle
	\begin{abstract}
Chorus detection is a challenging problem in musical signal processing as the chorus often repeats more than once in popular songs, usually with rich instruments and complex rhythm forms. Most of the existing works focus on the receptiveness of chorus sections based on some explicit features such as loudness and occurrence frequency. These pre-assumptions for chorus limit the generalization capacity of these methods, causing misdetection on other repeated sections such as verse. To solve the problem, in this paper we propose an end-to-end chorus detection model \emph{DeepChorus}, reducing the engineering effort and the need for prior knowledge. The proposed model includes two main structures: \textbf{i)} a Multi-Scale Network to derive preliminary representations of chorus segments, and \textbf{ii)} a Self-Attention Convolution Network to further process the features into probability curves representing chorus presence. To obtain the final results, we apply an adaptive threshold to binarize the original curve. The experimental results show that \emph{DeepChorus} outperforms existing state-of-the-art methods in most cases.

	\end{abstract}
	\begin{keywords}
		Chorus detection, Self-Attention, Multi-Scale Network, Music structure analysis, Music information retrieval
		\vspace{-.2cm}
	\end{keywords}

	\section{Introduction}
	\label{sec:intro}
	
    Music songs are highly structured, especially for popular music \cite{muller2012robust}. Among the sections in a pop song such as intro, verse, chorus, and bridge, the chorus can always best reflect the ``most catchy" section of a piece of music \cite{van2013analysis}. Chorus detection is such a task that aims to find a short, continuous segment of a piece of music that can nicely represent the whole piece. In music information retrieval (MIR) field, the automatic detection of the chorus is a challenging problem, while it has extensive application value in various scenarios, such as extracting audition for commercial, and helping the user to quickly and efficiently preview selections from a large music database \cite{yeh2007extraction}. 


Many previous studies have assumed that the melody patterns with the most repetition correspond to the chorus sections. Based on the assumption, methods with manual features or traditional machine learning have been attempted for chorus detection: two approaches for finding ``key phrases" based on clustering or hidden Markov model (HMM) was proposed in  \cite{logan2000music}; the other method considered the audio signal as several ``state" parts, using k-means and HMM to derive different sections automatically \cite{peeters2002toward}. Self-similarity matrix (SSM) is also widely used \cite{7340798, cooper2002automatic}, by which some significant change points can be derived and therefore the most-repeated sections are detected as the chorus. As a related research field, more research focused on generating ``thumbnails" from music songs, which is also known as ``thumbnailing". A chroma-based thumbnailing framework was proposed in \cite{bartsch2001catch}. A distance matrix representation was used \cite{eronen2007chorus} to produce music thumbnails. \cite{muller2012robust} proposed a fitness value that expresses the repetitiveness of the segments. In addition, many studies on music structure analysis involving the chorus sections also achieved certain results \cite{nieto2013convex, mcfee2014analyzing}.
	
With the development of deep learning in these years, the attempts of using deep neural networks (DNN) for chorus detection began to increase. 
\cite{ha2017automatic} and \cite{huang2018pop} used music genre/emotion classification as surrogate tasks to establish unsupervised models for music highlight extraction. \cite{shibata2020music} proposed a hybrid generative model for music
structure analysis, in which a recurrent neural network (RNN) is used for pre-processing. However, the lack of datasets with structure annotation limited the development of the supervised method for chorus detection. It is only in recent years that more audio datasets are annotated into chorus sections \cite{nieto2019harmonix}, alleviating the problem of insufficient training data. Using these datasets, \cite{wang2021supervised} proposed a supervised multi-task approach to detecting chorus in popular music. Combining boundary detection, the method demonstrated a great improvement in F-measure.

Though many studies have tried the DNN-based approach for chorus detection, the existing methods either rely on some surrogate tasks \cite{ha2017automatic, huang2018pop}, or introduce some additional information \cite{shibata2020music, wang2021supervised}. Most of these methods require complex post-processing. However, with the excessive involvement of hand-craft engineering, generalization ability may be hindered in such models. 

In this paper we propose \textit{DeepChorus}, an end-to-end chorus detection system based on a hybrid network of multi-scale convolution and self-attention, 
differentiating and extracting choruses by learning the latent features (such as harmonics from different instruments, or complex performance in low frequency caused by various rhythm patterns) of musical sections. Highlighting the similarity among different sections by learnable parameters, the required engineering effort and the need for prior knowledge can be reduced. To extract embedding vectors suitable for these characteristics from input Mel-spectrograms, a \emph{Multi-Scale Network} is introduced to focus on the latent chorus information both in high and low resolution. 
For further processing to emphasize and summarize the similarity of chorus sections, self-attention layers combined with convolutional layers are used in the \emph{Self-Attention Convolution Network} to compute the correlation between each vector. Moreover, 
the binarization process to obtain final chorus section results is implemented by adaptive thresholds and requires no prior knowledge, meaning that \emph{DeepChorus} can work end-to-end from input-output pairs. Experimental results on three public datasets show that \emph{DeepChorus} outperforms existing state-of-the-art methods in most cases.
    \begin{figure*}[t]
    \centering
    \includegraphics[width=17.8cm, trim={.9cm .5cm .4cm .7cm}, clip]{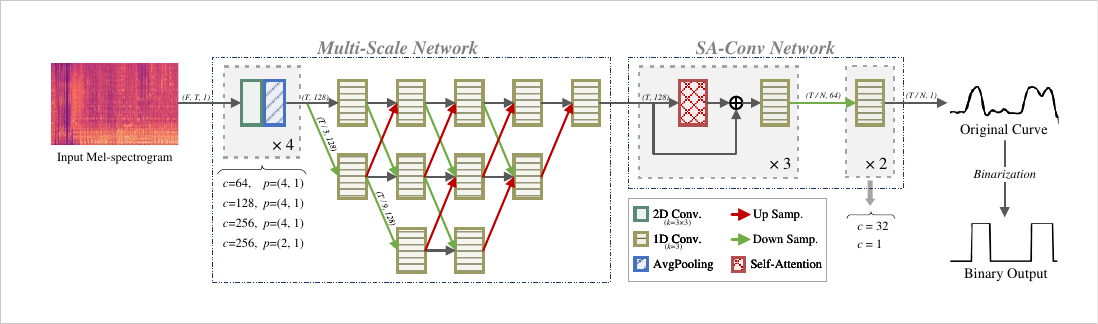}
    \caption{The overall architecture of \textit{DeepChorus}. $(F, T, C)$ denotes a feature map with length $F, T$ and $C$ on frequency axis, time axis, and channel axis respectively. $k$, $c$ and $p$ denote kernel size, channel number and pooling size respectively. Downsampling / upsampling are implemented by 1D  convolutional / transposed convolutional layers with strides of the same size as kernels.}
    \label{fig:backbone}
    \end{figure*} 

\section{Proposed Method}

The overall framework of \textit{DeepChorus} is illustrated in Fig. \ref{fig:backbone}. As shown in the figure, 
the output of the model is a binary vector, representing chorus or non-chorus. Because the Mel-scaled frequency matches closely the human auditory perception, the model is designed to take Mel-spectrograms as inputs. Note that the input song can be of arbitrary length, leading to a flexible model. In the rest of this section, we detail the main modules of \textit{DeepChorus}.
	
\subsection{Multi-Scale Network}
\label{sec:MS}

\begin{figure}[t]
    \vspace{-5pt}
    \subcaptionbox*{}{
    \vspace{-0.43cm}
    \includegraphics[width=.48\columnwidth]{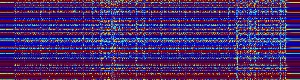}
    \hspace{-2pt}
    \includegraphics[width=.52\columnwidth]{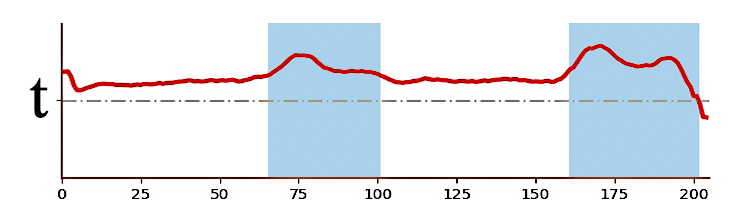}
}

    \subcaptionbox*{}{
    \includegraphics[width=.48\columnwidth]{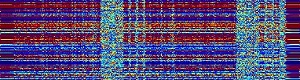}
    \hspace{-2pt}
    \includegraphics[width=.52\columnwidth]{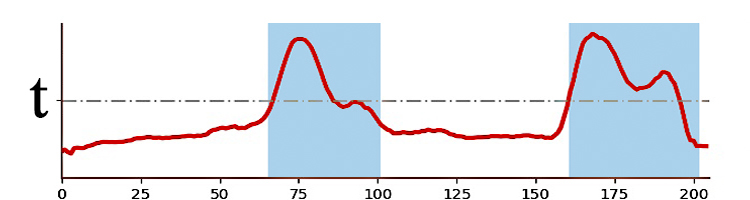}
}
\caption{Features after several layers (left) and the corresponding output curves (right) for ``Tenshi no utatane" in \texttt{RWC}, derived by a single-scale (upper) and a multi-scale (bottom) CNN model respectively. ``t" is the adaptive threshold and blue regions indicate the ground truth.}
\label{fig:vis_multi}
\end{figure}

For chorus detection, global information is particularly important for detecting homogeneity and repetitions, while local information is essential for finding an accurate position of change points. Considering the efficiency and flexibility of convolutional layers in MIR field \cite{huang2018pop, wang2021supervised}, we adopt a CNN-based architecture as the backbone. 
However, kernels are specifically designed to capture short-range information, 
a single convolutional layer cannot model dependencies that extend beyond the receptive field, causing unsatisfied results, and stacking multiple convolutional layers lead to an excess of parameters. 
The visualization of feature maps and output curves (as shown in Fig. \ref{fig:vis_multi}) also demonstrate convincingly that a traditional CNN structure is not such a suitable design for capturing global information for chorus detection. As a simple solution, downsampling operation is introduced in some network structures for segmentation, known as U-Net \cite{ronneberger2015u}. The dilated convolution \cite{yu2015multi} is also considered as an alternative to gain larger receptive fields. However, these solutions lead to the loss of details to some extent, which are important for locating chorus boundaries accurately. 

In order to tackle these issues, we introduce a multi-scale resolution network to our framework, which was originally designed for computer vision tasks \cite{wang2020deep}. The key idea of the presented strategy is: the input feature should be first downsampled to low resolution to facilitate the extraction of global information, and then merged to the high resolution. By downsampling / upsampling the feature to different scales and exchanging information repeatedly, embedding vectors capable of distinguishing chorus from non-chorus can be derived and region information can be highlighted for further processing after just a few layers.

The architecture of the Multi-Scale Network is composed of several branches for different scales. To convert the spectral information into the channel domain, we first process the spectrogram input $(F\times T\times 1)$ into a sequence with the shape of $(T, 128)$ by 2D convolutional layers with average pooling. Then in each branch, the features are rescaled over the time axis using 1D convolutional layers and transposed convolutional layers \cite{dumoulin2016guide}. 
The Multi-Scale Network conducts fusion repeatedly by concatenation along channel dimension such that representations in each branch receive information from different resolutions, leading to embedding vectors expressing chorus structure clearly. 

Fig. \ref{fig:vis_multi} shows the visualization of the output features of a traditional CNN without multi-scale structure and the Multi-Scale Network in \emph{DeepChorus}. The original curve before binarization is also shown in the figure. Note that the two models have a similar number of parameters. It is demonstrated that \emph{DeepChorus} with multi-scale structure obtains a clearer representation of chorus sections, while the output feature of traditional CNN network lacks obvious boundaries. The output curve of \emph{DeepChorus} also shows a more obvious change trend along the time axis, proving that the multi-scale structure is more suitable for chorus detection.










\subsection{Self-Attention Convolution Network}
\label{subsec:SA-Conv}

Success has been made in traditional methods by using SSM to detect chorus sections based on the homogeneity and repetitiveness of acoustic features \cite{muller2012robust}. However, a drawback in such methods is that additional hand-craft rules are often needed to form segments \cite{7340798, cooper2002automatic}, which leads to unsatisfying of robustness. 
Instead, we argue that the self-attention layer \cite{subakan2021attention} is a suitable alternative because of its strong ability to build pairwise relations using learnable parameters. Besides, embodying the principle of local processing, the convolutional layer can learn the local spatial features such as music structure edges and instrumental details within a segment. Adopting convolutional layers to further process the sequence helps to highlight the chorus boundary clearer. Therefore, we present a Self-Attention Convolution (SA-Conv) Network stacking self-attention layers and convolutional layers alternatively for the further processing of embedding sequence produced by the Multi-Scale Network.

We design an SA-Conv Block as the basic module. In the block, a self-attention layer and a convolutional layer are used. Three SA-Conv Blocks are stacked successively to form the main structure. Then a 1D convolutional layer whose kernel length and stride are $N$ is followed to downsample sequence length to $1/N$, where $N$ denotes the number of frames per second (fps) in the input Mel-spectrograms (in our experiments, $N$ is 43). Finally, two convolutional layers are adopted to process the sequence into a probability curve, representing the presence of chorus.

For an input sequence with $n$ elements $\bm{x} = (x_1,...,x_n), x_i\in \mathbb{R}^{d_x}$, the self-attention layer produces a new sequence $\bm{s} = (s_1,...,s_n)$ of the same length where $s_i\in \mathbb{R}^{d_s}$, and $\bm{s}$ is fused with $\bm{x}$ by concatenation. Each element of $\bm{s}$ is computed as a weighted sum of the linearly transformed input elements, and the results are fed to a softmax computation:
\begin{equation}
s_i=softmax\left [  \sum_{k=1}^{n} (p_{ik})(x_kW^V)\right ]
\label{eq1}
\end{equation}
where $p_{ik}$ is computed using a scaled dot-product to measure the similarities between $x_i$ and $x_k$:
\begin{equation}
    p_{ik}=\frac{(x_iW^Q)(x_kW^K)^T}{\sqrt{d_s}} 
    \label{eq2}
\end{equation}
$W^V, W^Q, W^K \in \mathbb{R}^{d_x\times d_s}$ are trainable parameter matrices in each layer. $\sqrt{d_s}=\parallel \bm{ss}^T-I\parallel ^2$ is a factor to prevent the result of dot-product being too large. Using scaled dot-product computation in Eq. \ref{eq2} enables linear transformations of the inputs of sufficient expressive power.

\begin{figure}[t]
    \centering
    \includegraphics[width=8.7cm, trim={0.8cm 0.8cm 0.5cm 0.6cm}, clip]{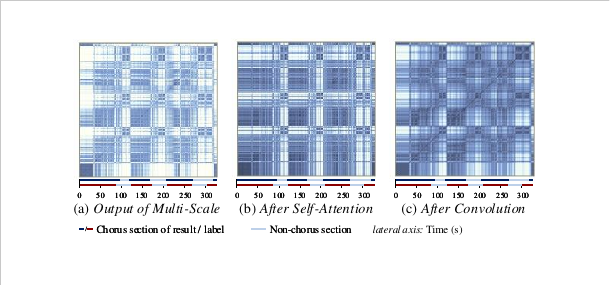}
    \caption{SSMs of SA-Conv Network of ``J-Song(Vagabond)" in \texttt{SL} dataset. Each point represents the distance between two vectors, and the darker the color, the closer the distance.
}
    \label{fig:SA-Conv_vis}
\end{figure}

To intuitively show the mechanism, we demonstrate the SSMs for the outputs of different layers in the SA-Conv Networks, as shown in Fig. \ref{fig:SA-Conv_vis}. It can be observed in Fig. \ref{fig:SA-Conv_vis}b that the self-attention layer summarizes the similarity to produce a more explicit representation, and therefore clusters the embeddings in the high-dimensional space for further processing. Furthermore, Fig. \ref{fig:SA-Conv_vis}c exhibits that the convolutional layer works as a supplement to the self-attention layer to enhance the local boundaries.

\subsection{Binarization}


In experiments we observed that impulse noise makes the resulted curves serrated, hindering the threshold selection for binarization. Therefore, we first apply a median filter  
to smooth the output curves. Supposing $\bm{y}=(y_1,...,y_n)$ is the original curve delivered by the SA-Conv Network, the smoothed curve $\bm{m}=(m_1,...,m_n)$ is given by:
\begin{equation}
   m_i=\left\{\begin{array}{ll}
   Med(y_{i-4},y_{i+4}),&{4<i\leq n-4},\\
   y_i,&{otherwise}.
   \label{MF}
\end{array}\right.
\end{equation}
where $Med(y_{i-4},y_{i+4})$ denotes the trimmed mean in the window. 

The probability curves for different songs are erratic and a fixed threshold cannot produce satisfactory results. To solve the problem, we apply adaptive thresholds for different curves to perform binarization. The adaptive threshold $t$ for $\bm{m}$ is defined by:
\begin{equation}
    t(\bm{m})=0.5 \times \left (m_{max} - m_{min} \right ) 
    \label{DT}
\end{equation}
where $m_{max}$ and $m_{min}$ correspond to the maximum and minimum of $\bm{m}$ respectively.





\section{Experiment}

\label{sec:exp}

\subsection{Experiment Setup}

\subsubsection{Datasets\& Metrics}
For training and validation, we select 886 tracks with ``chorus" label from \texttt{Harmonix} \cite{nieto2019harmonix} and 102 tracks by The Beatles and Michael Jackson from \texttt{Isophonics} \cite{isophonics} and regard ``refrain" labels as ``chorus".
Finally, a total of 988 tracks are used, among which 20 tracks are randomly chosen for validation, and the rest 968 tracks are adopted for training.

For testing, we used following public datasets: a) \texttt{RWC}: 100 popular songs from \cite{goto2002rwc} with structural labels provided by \cite{goto2006aist}; b) \texttt{SP}: 210 songs tagged as ``Popular" from SALAMI \cite{smith2011design}; and c) \texttt{SL}: 198 songs tagged as ``Live" class from SALAMI. Tracks are carefully selected to ensure that there is no overlap between \texttt{SP} and \texttt{SL}.

To compare with the latest work \cite{wang2021supervised}, F1 score and the area under curve (AUC) are chosen as evaluation metrics in this paper. We use AUC \cite{provost1997analysis} 
to study the model's ability to produce sensitive probability curves, while the F1 metric is a standard measurement for music structure analysis \cite{nieto2020audio}. 
For our ablation study, we supplement recall rate and precision rate as additional metrics, which are widely used to evaluate the effectiveness of information retrieval systems \cite{nieto2020audio}. 

\subsubsection{Training Details}{
To extract the input Mel-spectrograms, we first resample the original audio data to 22,050 Hz, then we use 512 overlapping windows of 2,048 frames and transform each window into a 128-band Mel-scaled magnitude spectrum. 
For training, the batch size we set is 4. In each epoch, 3,096 consecutive frames of each track (roughly 72 seconds long) are randomly selected for training. We choose the mean squared error as the loss function, and an Adam optimizer \cite{kingma2014adam} is applied with a learning rate of 0.0001. We evaluate F1 on the validation set every 50 iterations and save the model with the highest F1 score. The training is not stopped until F1 score has not improved for 500 iterations. 
}
    \begin{table}[t]
        \renewcommand\tabcolsep{2.8pt}
        \renewcommand\arraystretch{1.2}
        
        \centering
            \begin{tabular}{lcc|cc|cc|cc}
            \toprule[1pt]
            &\multicolumn{2}{c|}{F1}&\multicolumn{2}{c|}{AUC}&\multicolumn{2}{c|}{Recall}&\multicolumn{2}{c}{Precision}\\
            \textbf{Method}& \texttt{RWC} & \texttt{SP}&\texttt{RWC}&\texttt{SP}&\texttt{RWC}&\texttt{SP}&\texttt{RWC}&\texttt{SP}\\\hline\hline
            \emph{DeepChorus}&\textbf{.675}& \textbf{.611}&\textbf{.842} & \textbf{.780}&.582&\textbf{.844}&\textbf{.873}&\textbf{.523}\\\hline
            \emph{w/o SA-Conv} &.469&.430&.743&.709&.689&.694&.408&.356\\
            \emph{w/o Multi-scale}&.593&.555&.833&.763&\textbf{.760}&.766&.576&.518\\
            \emph{Fully CNN} &.546&.541&.693&.689&.739&.799&.507&.472\\
            \bottomrule
            \end{tabular}
    \caption{Results of Ablation Study on \texttt{RWC} and \texttt{SP} dataset, ``\emph{w/o SA-Conv.}", and ``\emph{w/o Multi-scale}" denote for without SA-Conv and Multi-Scale Network respectively. ``\emph{Fully CNN}" stands for a traditional single-scale fully convolution network.}
    \label{tab:compMulti}
  
\end{table}
\subsection{Ablation Study}
{
    \label{sec:AS}

To demonstrate how much the proposed structures contribute to the model, we conduct a set of experiments. Note that all the models in this section experiment have similar parameters (about 5M).

First, we replace the SA-Conv Network with a set of convolutional layers to justify the effectiveness of the SA-Conv Network. As shown in Table \ref{tab:compMulti}, the performances on \texttt{RWC} and \texttt{SP} datasets are both dropped to a large extent. When focusing on F1 metric, the model without the SA-Conv architecture shows significantly worse performance, decreased by 2.6\% on \texttt{RWC} and by 18.1\% on \texttt{SP}. Then, we remove the Multi-Scale Network and only keep the SA-Conv Network, using only ``single-scale" convolutional layers for producing embedding vectors. The performance on the two datasets is decreased by 8.2\% and by 5.6\% respectively. For remaining metrics, \emph{DeepChorus} also performs significantly better than other models in general. It is noteworthy that as shown in Fig. \ref{fig:vis_multi}, the original probability curves of the single-scale model (\emph{w/o Multi-scale}) are much flatter and mostly above the adaptive thresholds, which leads to a high recall rate but poor performance on AUC and precision.

To investigate the overall effectiveness of \emph{DeepChorus} structure, we remove both the SA-Conv Network and the Multi-Scale Network, and a traditional fully convolution network is adopted. The results on F1 are decreased by 11.3\% and 7.7\% on the two datasets respectively. For AUC, recall rate, and precision rate, there is also a certain degree of decline. The results indicate the obvious effect of the SA-Conv and Multi-Scale Network in \emph{DeepChorus}, proving them effective structures to detect the chorus more accurately.
}
\begin{table}[t]
    \centering
    \renewcommand\arraystretch{1}
    \begin{tabular}{l|cc|cc|cc}
        \toprule[1pt]
        &\multicolumn{2}{c|}{\texttt{RWC}}&\multicolumn{2}{c|}{\texttt{SP}}&\multicolumn{2}{c}{\texttt{SL}}\\
        \textbf{Method}&AUC&F1& AUC&F1& AUC&F1\\\hline\hline
        \emph{DeepChorus}             &\textbf{.842}&\textbf{.675}&\textbf{.780}&\textbf{.611}&\textbf{.765}&\textbf{.501}\\\hline
        \textit{Multi2021} \cite{wang2021supervised}       &.819&.643&.675&.473&.633&.380\\
        \textit{Highlighter} \cite{huang2018pop}      &.804&.407&.703&.303&.671&.251\\
        \textit{SCluster$^{*}$} \cite{mcfee2014analyzing}               &.533&.427&.545&.448&.551&.392\\
        \textit{CNMF$^{*}$} \cite{nieto2013convex}                   &.526&.403&.543&.422&.478&.340\\
        \bottomrule
    
    \end{tabular}
    \caption{Results of the proposed and baseline methods on \texttt{RWC}, \texttt{SP}, and \texttt{SL} dataset. Traditional methods are marked with asterisks (*).}
    \label{tab:comp}
    \vspace{-0.3cm}
\end{table}

\subsection{Comparison with Previous Work}
\begin{figure}
    \centering
    \includegraphics[width=8.5cm, trim={0.5cm 0.8cm 2.5cm 1.3cm}, clip]{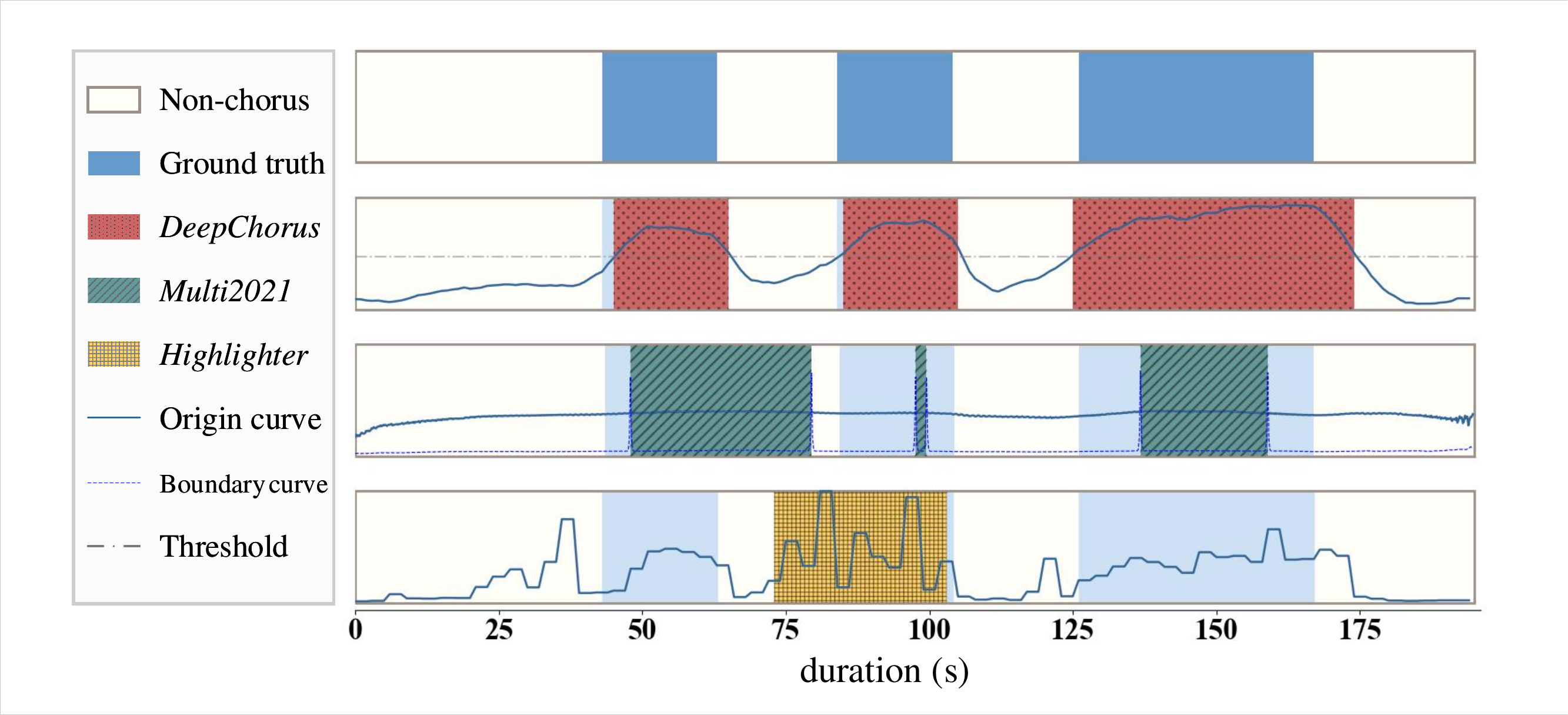}
    \caption{Chorus detection results for ``The Spy" in \texttt{SP} dataset using different methods.} 
    \label{fig:comp}
    \vspace{-0.5cm}
\end{figure}

{

To compare with previous work, we choose four models as baseline methods: \emph{CNMF} \cite{nieto2013convex}, \emph{SCluster} \cite{mcfee2014analyzing}, \emph{Highlighter} \cite{huang2018pop}, and \emph{Multi2021} \cite{wang2021supervised}. Among them, \emph{CNMF} and \emph{SCluster} are traditional methods, and the implementation by MSAF \cite{nieto2016systematic} (a framework contains various algorithms for music structure analysis) is used. For \emph{Highlighter}, an unsupervised CNN-based model, the pre-trained model provided by the author is used for evaluation. \emph{Multi2021} is a supervised CNN-based model with state-of-the-art performance among the chosen systems. For a fair comparison, we train the model on our training datasets. The model consists of nearly 7M parameters. Due to the limitations of experimental conditions, the batch size 256 in the original experiment is not achievable, and therefore we set batch to 32 as an alternative (roughly 600 seconds per mini-batch, about twice as long as ours).

The experimental results on the three test datasets are shown in Table \ref{tab:comp}
. The results clearly confirm that \emph{DeepChorus} outperforms four baseline methods significantly for both AUC and F1
. When focusing on the \texttt{RWC} dataset, which is widely used in music structure analysis, \emph{DeepChorus} exhibits an improved performance by 2.5\% higher than the other methods on F1 on average. For \emph{Multi2021}, which we consider as the state-of-the-art in chorus detection, \emph{DeepChorus} outperforms it by 8.67\% on AUC and by 9.7\% on F1 on average. This gap may be largely due to the heavy dependency on prior information. For example, an average number of chorus is used as prior knowledge in \emph{Multi2021}, which can be considered as a reason for the large performance gap between \emph{Deepchorus} and \emph{Multi2021} on \texttt{SP}. Because in \texttt{SP}, the repeat times of chorus varies notably among. 
The \texttt{SL} dataset is made up of live performance tracks, which leads to the degraded quality of audio signals, making chorus detection more difficult compared to the other datasets but also closer to the real-world scenarios.

To investigate how \emph{DeepChorus} improves the performance in chorus detection, a case study is performed on a randomly selected song. We choose \emph{Multi2021} and \emph{Highlighter} to compare with because both of them are DNN-based methods. As depicted in Fig. \ref{fig:comp}, we can observe that there are fewer misdetections on chorus sections and more accurate boundaries for \emph{DeepChorus}. From the visualization result of \emph{Multi2021}, we can find that its original curves are too flat that nearly all values are below the fixed threshold 0.5, and therefore a boundary curve is indispensable to provide additional information for chorus section. Compared with \emph{Highlighter}, we can say that the performance gains of \emph{DeepChorus} can be attributed to the ability to detect multiple chorus sections at a time and to locate them with better precision. 
}

\section{Conclusion}

In this paper, we have proposed \emph{DeepChorus}, an end-to-end system for chorus detection, mainly consisting of a Multi-Scale Network and a Self-Attention Convolution Network. Trained on mass data, latent chorus information in Mel-spectrograms can be learned to differentiate and highlight choruses, and therefore the model gains higher generalization capacity. Experimental results show \emph{DeepChorus} outperforms the existing state-of-the-art methods on three datasets significantly. Designing a method to improve the performance of chorus detection remains an avenue for future work.

The code of 
\emph{DeepChorus} and the reproduced model is available at \texttt{https://github.com/Qqi-HE/DeepChorus}.

\section{Acknowledgement}
    \vspace{-.05cm}	
    This work was supported by National Key R\&D Program of China (2019YFC1711800) and NSFC (62171138).
    \vspace{-.25cm}	
	\bibliographystyle{IEEEbib}
	\bibliography{DeepChorus}
\end{document}